\documentclass[12pt]{article}
\usepackage{amsmath}
\usepackage{amssymb}
\usepackage{enumerate}

\begin{document}
\title{On the structure of optimal sets for tensor product channel.}
\author{M.E.Shirokov \thanks{Steklov Mathematical Institute, 119991 Moscow,
Russia}}
\date{}
\maketitle

\section{Introduction}

The minimal output entropy and the Holevo capacity are important
characteristics of a quantum channel. By using properties of
optimal ensembles \cite{Sch-West-1} and the convex duality
observations \cite{A&B} it is possible to show that the both this
quantities can be expressed via the conjugate function to the
output entropy of a channel (proposition 1). The convex duality
approach also provides the characterization of the average state
of an optimal ensemble for an arbitrary quantum channel
(proposition 2).

For a given quantum channel we consider two \textit{optimal} sets
of states, related to the Holevo capacity and to the minimal
output entropy of this channel correspondingly. Some properties of
these sets as well as the necessary and sufficient condition for
their coincidence are obtained (propositions 3,4, corollary 1).

One of the most interesting open problems in the quantum information
theory is the additivity conjecture for the Holevo capacity of a
quantum channel. The role of this conjecture was stressed by Shor's
recent proof of its equivalence to several other (sub)additivity
problems, in particular, to the additivity conjecture for the
minimal output entropy. The relations between additivity properties
for two quantum channels and the structure of the optimal sets for
tensor product of these channels are considered (propositions 5,6).
It turns out that these additivity properties are connected with the
special \textit{hereditary} property of the optimal sets for tensor
product channel. This hereditary property seems to be of independent
interest and imposes strong restrictions on the structure of a set
of states having this property (the theorem in section 4). This
leads to some observations concerning the structure of the optimal
sets for tensor product channels. We explore the structural
properties of these optimal sets under the two different
assumptions.

The first assumption means that for tensor product of two channels
with arbitrary constraints there exists optimal ensemble with the
product state average. This assumption implies neither additivity of
the minimal output entropy nor additivity of the Holevo capacity,
but it can be considered as the first step in their proving
\cite{MSW}. It turns out that exactly this assumption guarantees
hereditary property of the both optimal sets for tensor product
channel (proposition 7). This and the theorem in section 4 provides
observations concerning relations between this assumption and the
additivity properties (corollaries 2-6).

The second assumption is stronger than the first one and means
additivity of the Holevo capacity for two channels with arbitrary
constraints. This assumption implies additivity of the minimal
output entropy and of the Holevo capacity. It turns out that exactly
is this assumption guarantees strong hereditary property of the both
optimal sets for tensor product channel and provides the natural
"projective" relations between these sets and the optimal sets for
single channels (proposition 8). These projective relations imply
strong restrictions on existence of entangled states in the optimal
sets for a tensor product channel (corollary 7 and the notes below).

\section{The case of single channel}

Let $\Phi :\mathfrak{S}(\mathcal{H})\mapsto
\mathfrak{S}(\mathcal{H}^{\prime })$ be an arbitrary channel,
where $\mathcal{H}$ and $\mathcal{H}^{\prime }$ are Hilbert spaces
of dimensions $d$ and $d^{\prime }$ correspondingly. The Holevo
quantity \cite{H-1} of an ensemble $\{\pi_{i},\rho_{i}\}$ of
states in $\mathfrak{S}(\mathcal{H})$ is defined by
\[
\chi _{\Phi}\left( \{\pi_{i},\rho_{i}\}\right)
=H\left(\sum_{i}\pi_{i}\Phi(\rho_{i})\right)
-\sum_{i}\pi_{i}H(\Phi(\rho_{i})),
\]
where $H(\Phi(\rho))=-\mathrm{Tr}\Phi(\rho)\log\Phi(\rho)$  is the
output entropy of the channel $\Phi$. The $\chi$-function of the
channel $\Phi$ is defined as
\[
\chi_{\Phi}(\rho)=\max_{\sum_{i}\pi_{i}\rho_{i}=\rho}\chi
_{\Phi}(\{\pi_{i},\rho_{i}\}).\vspace{10pt}
\]
Let $\mathcal{A}$ be an arbitrary closed and convex subset of
$\mathfrak{S}(\mathcal{H})$. The channel $\Phi$ with the
constraint on input ensemble $\{\pi_{i},\rho_{i}\}$ defined by the
inclusion $\bar{\rho}\in\mathcal{A}$ is called the
$\mathcal{A}$-constrained channel $\Phi$. The Holevo capacity of
the $\mathcal{A}$-constrained channel $\Phi$ can be defined as
\[
\bar{C}(\Phi;\mathcal{A})=\max_{\rho\in\mathcal{A}}\chi_{\Phi}(\rho).
\]

The ensemble $\{\pi_{i},\rho_{i}\}$ with the average
$\bar{\rho}=\sum_i\pi_{i}\rho_{i}$ is called \textit{optimal} for
the $\mathcal{A}$-constrained channel $\Phi$ if
\[
\chi _{\Phi}\left( \{\pi_{i},\rho_{i}\}\right)=\chi _{\Phi}\left(
\bar{\rho}\right)=\bar{C}(\Phi;\mathcal{A}).
\]
The average state of this ensemble is called \textit{optimal
average state} for the $\mathcal{A}$-constrained channel $\Phi$.

It is known that while the optimal ensemble may be not unique, its
average $\bar{\rho}$ has the unique image $\Phi(\bar{\rho})$ (it
follows, for example, from corollary 1 in \cite{H-Sh-1}). This
unique state in $\mathfrak{S}(\mathcal{H}')$ will be denoted as
$\Omega(\Phi;\mathcal{A})$.

Let $\bar{C}(\Phi)=\bar{C}(\Phi;\mathfrak{S}(\mathcal{H}))$ and
$\Omega(\Phi)=\Omega(\Phi;\mathfrak{S}(\mathcal{H}))$ be the
Holevo capacity and the image of the optimal average state
correspondingly for the unconstrained channel $\Phi$.

The minimal output entropy of the channel $\Phi$ is defined by
\[
H_{\mathrm{min}}(\Phi)=\min_{\rho \in
\mathfrak{S}(\mathcal{H})}H(\Phi(\rho)).
\]
The state $\rho$ is called \textit{optimal for the channel} $\Phi$
if
\[
H(\Phi(\rho))=H_{\mathrm{min}}(\Phi).
\]

Consider the convex closure $\hat{H}_{\Phi}(\rho)$ of the function
$H(\Phi(\rho))$, defined by
$$
\hat{H}_{\Phi}(\rho)=\min\sum_{i}\pi_{i}H(\Phi(\rho_{i})),
$$
where the minimum is over all ensembles $\{\pi_{i},\rho_{i}\}$  of
pure state with the average $\rho$. By definition we have
\begin{equation}\label{chi-rep}
\chi_{\Phi}(\rho)=H(\Phi(\rho))-\hat{H}_{\Phi}(\rho)).
\end{equation}

Note that $\bar{C}(\Phi)$ is the maximal gap between
$H(\Phi(\rho))$ and $\hat{H}_{\Phi}(\rho )$.

Consider the conjugate function \cite{A&B},\cite{JT}
\begin{equation}\label{H*-def}
H_{\Phi}^{*}(A)=\max\limits_{\rho\in\mathrm{Extr}\mathfrak{S}(\mathcal{H})}
\left(\mathrm{Tr}A\rho-H(\Phi(\rho))\right).
\end{equation}
By duality
\begin{equation}\label{c-d-1}
\hat{H}_{\Phi}(\rho)=\max\limits_{A\in\mathfrak{B}_{h}(\mathcal{H})}
\left(\mathrm{Tr}A\rho-H_{\Phi}^{*}(A)\right)
\end{equation}
and
\begin{equation}\label{c-d-2}
H_{\Phi}^{*}(A)=\max\limits_{\rho\in\mathfrak{S}(\mathcal{H})}
\left(\mathrm{Tr}A\rho-\hat{H}_{\Phi}(\rho)\right).
\end{equation}

\textbf{Proposition 1.} \textit{The minimal output entropy}
$H_{\mathrm{min}}(\Phi)$ \textit{and the Holevo capacity}
$\bar{C}(\Phi)$ \textit{of the channel} $\Phi$ \textit{can be
defined in terms of the conjugate function} $H_{\Phi}^{*}$
\textit{by the expressions}\footnote{it is assumed that the
operator $\log\Phi(\rho)$ has the same null space as
$\Phi(\rho)$.}
$$
\begin{array}{c}
H_{\mathrm{min}}(\Phi)=\min\limits_{\rho\in\mathfrak{S}(\mathcal{H})}\hat{H}_{\Phi}(\rho)
=\lambda-H_{\Phi}^{*}(\lambda
\mathcal{I}_{\mathcal{H}}),\quad\forall\lambda\in\mathbb{R},\\\\
\bar{C}(\Phi)=\min\limits_{\rho\in\mathfrak{S}(\mathcal{H})}
H_{\Phi}^{*}(\Phi^{*}(-\log\Phi(\rho)))=
H_{\Phi}^{*}(\Phi^{*}(-\log\Omega(\Phi))).
\end{array}
$$
\textit{where $\Phi^*$ is the dual map for the channel} $\Phi$.

\textbf{Proof.} The expression for $H_{\mathrm{min}}$ is obvious.

From the definition (\ref{H*-def}) we have
$$
H_{\Phi}^{*}(\Phi^{*}(-\log\Phi(\rho))=
\max\limits_{\omega\in\mathrm{Extr}(\mathfrak{S}(\mathcal{H}))}H(\Phi(\omega)\|\Phi(\rho)),
$$
where $H(\cdot||\cdot)$ is the relative entropy \cite{Sch-West-2}.
Each state in $\mathfrak{S}(\mathcal{H})$ can be represented as a
convex combination of some states in
$\mathrm{Extr}\mathfrak{S}(\mathcal{H})$. So, the convexity of the
quantum relative entropy implies
$$
\max\limits_{\omega\in\mathrm{Extr}\mathfrak{S}(\mathcal{H})}H(\Phi(\omega)\|\Phi(\rho))=
\max\limits_{\omega\in\mathfrak{S}(\mathcal{H})}H(\Phi(\omega)\|\Phi(\rho)).
$$

The statement of the proposition follows from the two last
equalities and expression (19) in \cite{Sch-West-1} for the Holevo
capacity.$\square$

By proposition 1 in \cite{A&B} if some operator
$A\in\mathfrak{B}_{h}(\mathcal{H})$ is optimal for a given state
$\rho\in\mathfrak{S}(\mathcal{H})$ in (\ref{c-d-1}) then this
$\rho$ is optimal for the operator $A$ in (\ref{c-d-2}) and vice
versa. In addition, each pure state $\rho_{i}$ in the optimal
decomposition of $\rho$ is optimal for the operator $A$ in
(\ref{H*-def}). In terms of convex analysis this means that $A$ is
contained in the subdifferential $\partial\hat{H}_{\Phi}(\rho)$ of
the function $\hat{H}$ at the point $\rho$ and  all $\rho_{i}$ as
well as $\rho$ are contained in the subdifferential $\partial
H_{\Phi}^{*}(A)$ of the function $H_{\Phi}^{*}$ at the point
$A$.\footnote{The subdifferential $\partial f$ of a convex
function $f$ at a point $x$ is the (convex) set of all linear
functionals $l$ such that the function $f-l$ has minimum at the
point $x$ \cite{JT}.}

\textbf{Proposition 2.} \textit{The state} $\rho$ \textit{is the
average state of the optimal ensemble for the channel} $\Phi$
\textit{if and only if the operator}  $\Phi^*(-\log\Phi(\rho))$
\textit{is optimal for the state} $\rho$ \textit{in (\ref{c-d-1})
(and, necessarily,  the state} $\rho$ \textit{is optimal for the
operator} $\Phi^*(-\log\Phi(\rho))$ \textit{in (\ref{c-d-2})). In
terms of the convex analysis $\rho$ is the average state of the
optimal ensemble if and only if}
$$
\Phi^*(-\log\Phi(\rho))\in\partial\hat{H}_{\Phi}(\rho).
$$

\textbf{Proof.} If the state $\rho$ is the average state of the
optimal ensemble then, by the maximal distance property
\cite{Sch-West-1}, we have
$$
\begin{array}{c}
H^{*}(-\log\Phi(\rho))=
\max\limits_{\omega\in\mathfrak{S}(\mathcal{H})}H(\Phi(\omega)\|\Phi(\rho))=
\chi_{\Phi}(\rho)\\=-\textup{Tr}\left[\Phi(\rho)\log\Phi(\rho)\right]-\hat{H}_{\Phi}(\rho)=
\textup{Tr}\left[\rho\Phi^{*}(-\log\Phi(\rho))\right]-\hat{H}_{\Phi}(\rho).
\end{array}
$$
In accordance with (\ref{c-d-2}) this means that the state $\rho$
is optimal for the operator $\Phi^{*}(-\log\Phi(\rho))$ and hence
$\Phi^{*}(-\log\Phi(\rho))\in\partial\hat{H}_{\Phi}(\rho)$.

If $\Phi^{*}(-\log\Phi(\rho))\in\partial\hat{H}_{\Phi}(\rho)$ we
have for any state $\omega$ in $\mathfrak{S}(\mathcal{H})$
$$
\begin{array}{c}
H(\Phi(\omega)\|\Phi(\rho))\leq\max\limits_{\omega'\in\mathfrak{S}(\mathcal{H})}
H(\Phi(\omega')\|\Phi(\rho))
=H_{\Phi}^{*}(\Phi^{*}(-\log\Phi(\rho)))\\=
\textup{Tr}\left[\Phi(\rho)(-\log\Phi(\rho))\right]-\hat{H}_{\Phi}(\rho)=
\chi_{\Phi}(\rho).
\end{array}
$$
By the maximal distance property this means that the state $\rho$
is the average state of the optimal ensemble.$\square$

Corollary 1 in \cite{H-Sh-1} implies the  following inequality
\begin{equation}\label{chi-ineq}
\chi _{\Phi}(\rho )+H(\Phi(\rho)\Vert\Omega(\Phi))\leq
\bar{C}(\Phi ),\quad \forall\rho\in\mathfrak{S}(\mathcal{H}).
\end{equation}

It turns out that the set of states for which equality holds in
the above inequality plays an important role in all questions
related to the Holevo capacity.

\textbf{Definition 1.} \textit{The subset $\mathcal{A}_{C}^{\Phi}$
of $\mathfrak{S}(\mathcal{H})$, consisting of all states $\rho$
such that
\[
\chi _{\Phi}(\rho )+H(\Phi(\rho)\Vert\Omega(\Phi))=\bar{C}(\Phi )
\]
is called the optimal set for the channel $\Phi$ in the sense of
the Holevo capacity}.

The set $\mathcal{A}_{C}^{\Phi}$ can be defined as a maximal
convex subset of $\mathfrak{S}(\mathcal{H})$, containing
$\bar{\rho}$ and having affine restriction of the function
$\hat{H}_{\Phi}$, simular to the analogous definition of the set
$\Omega_{\omega}$ in \cite{Uhl}. This observation can be deduced
from the remark before proposition 2 and the following
proposition, which justifies the term "optimal set" for
$\mathcal{A}_{C}^{\Phi}$.

\textbf{Proposition 3.} \textit{The set} $\mathcal{A}_{C}^{\Phi}$
\textit{is a closed convex subset of} $\mathfrak{S}(\mathcal{H})$,
\textit{coinciding with the subdifferential} $\partial
H_{\Phi}^{*}(\Phi^{*}(-\log\Omega(\Phi)))$ \textit{of the
function} $H_{\Phi}^{*}$ \textit{at the point}
$\Phi^{*}(-\log\Omega(\Phi))$\textit{. The set of extreme points
of $\mathcal{A}_{C}^{\Phi}$ consists of all pure states} $\rho$
\textit{such that} $H(\Phi(\rho)\Vert\Omega(\Phi))=\bar{C}(\Phi)$
\textit{and, hence, contains elements of any pure state optimal
ensemble for the channel} $\Phi$.\footnote{The assertion that any
extreme point of $\mathcal{A}_{C}^{\Phi}$ can be included
(nontrivially) in some optimal ensemble is an interesting open
problem. It is true for some special channels, in particular, for
covariant channels (see below).}

\textbf{Proof.} By proposition 1 for any
$\rho\in\mathcal{A}_{C}^{\Phi}$ we have
$$
\begin{array}{c}
 H_{\Phi}^{*}(\Phi^{*}(-\log\Omega(\Phi)))=\bar{C}(\Phi)=\chi
_{\Phi}(\rho )+H(\Phi(\rho) \Vert
\Omega(\Phi))\\\\=\mathrm{Tr}\rho\Phi^{*}(-\log\Omega(\Phi))-\hat{H}_{\Phi}(\rho)
\end{array}
$$
By the remark before proposition 2 this means that $\rho$ lies in
the subdifferential $\partial
H_{\Phi}^{*}(\Phi^{*}(-\log\Omega(\Phi)))$ of the function
$H_{\Phi}^{*}$ at the point $\Phi^{*}(-\log\Omega(\Phi))$.

By the properties of subdifferential the set
$\mathcal{A}_{C}^{\Phi}=\partial
H_{\Phi}^{*}(\Phi^{*}(-\log\Omega(\Phi)))$ is convex and closed.
This also can be directly deduced from definition 1 and inequality
(\ref{chi-ineq}).

The purity of any extreme point of $\mathcal{A}_{C}^{\Phi}$ is
proved as follows. If $\chi_{\Phi}(\rho)>0$ for a state $\rho$ in
$\mathcal{A}_{C}^{\Phi}$ then $H(\Phi(\rho))>\hat{H}_{\Phi}(\rho)$
and, so, there exists a nontrivial decomposition of $\rho$ optimal
in the sense of the definition of $\hat{H}_{\Phi}(\rho)$. By the
remark  before proposition 2 all elements of this decomposition
lie in $\mathcal{A}_{C}^{\Phi}=\partial
H_{\Phi}^{*}(\Phi^{*}(-\log\Phi(\bar{\rho})))$. Hence the above
state $\rho$ is not an extreme point of $\mathcal{A}_{C}^{\Phi}$
and, so, the function $\chi_{\Phi}$ equals to zero on the set of
all extreme point of $\mathcal{A}_{C}^{\Phi}$. By the strong
concavity of the quantum entropy the equality
$\chi_{\Phi}(\rho)=0$ for some mixed state
$\rho\in\mathcal{A}_{C}^{\Phi}$ implies that the channel $\Phi$
has the same output for any pure state decomposition of $\rho$.
Hence all elements of this decomposition are also in
$\mathcal{A}_{C}^{\Phi}$ and, so, the above state $\rho$ is not an
extreme point of $\mathcal{A}_{C}^{\Phi}$. In other words, all
extreme points of $\mathcal{A}_{C}^{\Phi}$ are pure states.

The assertion concerning optimal ensemble follows from the maximal
distance property \cite{Sch-West-1}. $\square$

The minimal output entropy provides another optimal set.

\textbf{Definition 2.} \textit{The convex closure
$\mathcal{A}_{E}^{\Phi}$ of all optimal states for the channel
$\Phi$ in $\mathfrak{S}(\mathcal{H})$ is called the optimal set
for the channel $\Phi$ in the sense of the minimal output
entropy.}

By proposition 1 the set $\mathcal{A}_{E}^{\Phi}$ coincides with
the subdifferential $\partial H_{\Phi}^{*}(\lambda
\mathcal{I}_{\mathcal{H}})$ of the function $H_{\Phi}^{*}$ at the
point $\lambda \mathcal{I}_{\mathcal{H}}$ for any $\lambda$.

We will use the following characterization of extreme points of
the above optimal sets $\mathcal{A}_{E}^{\Phi}$ and
$\mathcal{A}_{C}^{\Phi}$.

\textbf{Lemma 1.} \textit{A pure state
$|\varphi\rangle\langle\varphi|$ is an extreme point of the set
$\mathcal{A}_{E}^{\Phi}$ if and only if
$$
\Phi^{*}(-\log\Phi(|\varphi\rangle\langle\varphi|))|\varphi\rangle=H_{\mathrm{min}}(\Phi)|\varphi\rangle.
$$
A pure state $|\varphi\rangle\langle\varphi|$ is an extreme point
of the set $\mathcal{A}_{C}^{\Phi}$ if and only if}
$$
\Phi^{*}(-\log\Omega(\Phi)+\log\Phi(|\varphi\rangle\langle\varphi|))|\varphi\rangle
=\bar{C}(\Phi)|\varphi\rangle.
$$

\textbf{Proof.} The both assertions of the lemma are reformulations
of the known results \cite{Gio},\cite{R&S}, which were obtained (by
author's knowledge) for the first time by P.W.Shor. We give their
simple proofs obtained independently and based on the properties of
the relative entropy.

By the definition a pure state $|\varphi\rangle\langle\varphi|$ is
an extreme point of the set $\mathcal{A}_{E}^{\Phi}$ if and only if
the vector $|\varphi\rangle$ is a minimum point of the function
$$
\mathcal{H}\ni|\psi\rangle\mapsto
H(\Phi(|\psi\rangle\langle\psi|))=\langle\psi|\Phi^{*}(-\log\Phi(|\psi\rangle\langle\psi|))|\psi\rangle
$$
and the corresponding minimum value of this function is
$H_{\mathrm{min}}(\Phi)$. This and nonnegativity of the relative
entropy imply
$$
\begin{array}{c}
H_{\mathrm{min}}(\Phi)=\langle\varphi|\Phi^{*}(-\log\Phi(|\varphi\rangle\langle\varphi|))|\varphi\rangle\\\\\leq
\langle\psi|\Phi^{*}(-\log\Phi(|\psi\rangle\langle\psi|))|\psi\rangle\leq
\langle\psi|\Phi^{*}(-\log\Phi(|\varphi\rangle\langle\varphi|))|\psi\rangle
\end{array}
$$
for arbitrary unit vector $|\psi\rangle$. It follows that
$|\varphi\rangle$ is an eigen vector of the hermitian operator
$\Phi^{*}(-\log\Phi(|\varphi\rangle\langle\varphi|))$
corresponding to its minimal eigen value $H_{\mathrm{min}}(\Phi)$.

To prove the second part of the lemma note that by proposition 3 a
pure state $|\varphi\rangle\langle\varphi|$ is an extreme point of
the set $\mathcal{A}_{C}^{\Phi}$ if and only if the vector
$|\varphi\rangle$ is a maximum point of the function
$$
\mathcal{H}\ni|\psi\rangle\mapsto
H(\Phi(|\psi\rangle\langle\psi|)\Vert
\Omega(\Phi))=\langle\psi|\Phi^{*}(-\log\Omega(\Phi))+\log\Phi(|\psi\rangle\langle\psi|))|\psi\rangle
$$
and the corresponding maximum value of this function is
$\bar{C}(\Phi)$. This and nonnegativity of the relative entropy
imply
$$
\begin{array}{c}
\bar{C}(\Phi)=\langle\varphi|\Phi^{*}(-\log\Omega(\Phi)+\log\Phi(|\varphi\rangle\langle\varphi|))|\varphi\rangle\\\\\geq
\langle\psi|\Phi^{*}(-\log\Omega(\Phi)+\log\Phi(|\psi\rangle\langle\psi|))|\psi\rangle\\\\\geq
\langle\psi|\Phi^{*}(-\log\Omega(\Phi)+\log\Phi(|\varphi\rangle\langle\varphi|))|\psi\rangle
\end{array}
$$
for arbitrary unit vector $|\psi\rangle$. It follows that
$|\varphi\rangle$ is an eigen vector of the hermitian operator
$\Phi^{*}(-\log\Omega(\Phi)+\log\Phi(|\varphi\rangle\langle\varphi|))$
corresponding to its maximal eigen value $\bar{C}(\Phi)$.

The converse statements of the both parts of this lemma obviously
follows from the definitions of the sets $\mathcal{A}_{E}^{\Phi}$
and $\mathcal{A}_{C}^{\Phi}$. $\square$

It is interesting to consider relations between the above optimal
sets $\mathcal{A}_{E}^{\Phi}$ and $\mathcal{A}_{C}^{\Phi}$ for
particular channels. The following proposition provides necessary
and sufficient condition for coincidence of these sets.

\textbf{Proposition 4.} \textit{Let
$\Phi:\mathfrak{S}(\mathcal{H})\mapsto
\mathfrak{S}(\mathcal{H}^{\prime })$ be an arbitrary channel with
the dual map $\Phi^{*}:\mathfrak{B}_{h}(\mathcal{H}')\mapsto
\mathfrak{B}_{h}(\mathcal{H})$. Let $P_{\Phi}$ be the projector on
the minimal subspace $\mathcal{H}_{\Phi}$ of the space
$\mathcal{H}$ containing supports of all states in
$\mathcal{A}_{E}^{\Phi}\cup\mathcal{A}_{C}^{\Phi}$.}

\textit{The coincidence
$\mathcal{A}_{E}^{\Phi}=\mathcal{A}_{C}^{\Phi}$ takes place if and
only if the following condition holds}\footnote{It follows from
the proof below that necessarily
$\lambda=\bar{C}(\Phi)+H_{\mathrm{min}}(\Phi)$ in (\ref{c-cond}).}
\begin{equation}\label{c-cond}
\exists\lambda\in\mathbb{R}:\Phi^{*}(-\log\Omega(\Phi))P_{\Phi}=\lambda
P_{\Phi}.
\end{equation}

\textbf{Proof.} Let equality (\ref{c-cond}) take place with some
$\lambda$. By the definition of the subspace $\mathcal{H}_{\Phi}$
for arbitrary state $\rho$ in
$\mathcal{A}_{E}^{\Phi}\cup\mathcal{A}_{C}^{\Phi}$ we have $\rho
P=\rho$ and hence
$\mathrm{Tr}\rho\Phi^{*}(-\log\Omega(\Phi))=\lambda$. Consider the
function
$$
\begin{array}{c}
\rho\mapsto H(\Phi(\rho)\|\Omega(\Phi))=
\mathrm{Tr}\rho\Phi^{*}(-\log\Omega(\Phi))-H(\Phi(\rho))=\lambda-H(\Phi(\rho))
\end{array}
$$
on the set $\mathcal{A}_{E}^{\Phi}\cup\mathcal{A}_{C}^{\Phi}$. By
proposition 3 a pure state $\rho$ lies in
$\mathrm{Extr}\mathcal{A}_{C}^{\Phi}$ if and only if the above
function achieves its maximum $\bar{C}(\Phi)$ at the state $\rho$.
This means that the function $H(\Phi(\rho))$ achieves at the state
$\rho$ its minimum on the set
$\mathcal{A}_{E}^{\Phi}\cup\mathcal{A}_{C}^{\Phi}$. By the
definition of the set $\mathcal{A}_{E}^{\Phi}$ this is valid if and
only if the state $\rho$ lies in $\mathcal{A}_{E}^{\Phi}$.

Conversely, let $\mathcal{A}_{E}^{\Phi}=\mathcal{A}_{C}^{\Phi}$,
then by the definition of the subspace $\mathcal{H}_{\Phi}$ there
exists a finite collection
$\{|\varphi_{i}\rangle\langle\varphi_{i}|\}_{i=1}^{n}$ of pure
states in $\mathcal{A}_{E}^{\Phi}=\mathcal{A}_{C}^{\Phi}$ such that
the vectors $\{|\varphi_{i}\rangle\}$ form a basis of
$\mathcal{H}_{\Phi}$.

By lemma 1
$$
\Phi^{*}(-\log\Omega(\Phi))|\varphi_{i}\rangle=
(\bar{C}(\Phi)+H_{\mathrm{min}}(\Phi))|\varphi_{i}\rangle
$$
for all $i$, which implies (\ref{c-cond}) with
$\lambda=\bar{C}(\Phi)+H_{\mathrm{min}}(\Phi)$.$\square$

Proposition 4 can be reformulated as follows.

\textbf{Corollary 1.} \textit{Let
$\Phi:\mathfrak{S}(\mathcal{H})\mapsto
\mathfrak{S}(\mathcal{H}^{\prime })$ be an arbitrary channel with
the dual map $\Phi^{*}:\mathfrak{B}_{h}(\mathcal{H}')\mapsto
\mathfrak{B}_{h}(\mathcal{H})$. The coincidence
$\mathcal{A}_{E}^{\Phi}=\mathcal{A}_{C}^{\Phi}$ takes place if the
following condition holds}
\begin{equation}\label{c-cond-2}
\exists\lambda\in\mathbb{R}:\Phi^{*}(-\log\Omega(\Phi))=\lambda
\mathcal{I}_{\mathcal{H}}. \end{equation}

\textit{The above condition is also necessary for the coincidence
$\mathcal{A}_{E}^{\Phi}=\mathcal{A}_{C}^{\Phi}$ if there exists a
full rank state in $\mathcal{A}_{E}^{\Phi}=\mathcal{A}_{C}^{\Phi}$.}

\textit{In the case of bistochastic channel $\Phi$ (for which
$\Phi(d^{-1}\mathcal{I}_{\mathcal{H}})=d^{\prime-1}\mathcal{I}_{\mathcal{H}'}$)
the condition (\ref{c-cond-2}) means that $\Omega(\Phi)$ is the
chaotic state $d^{\prime -1}\mathcal{I}_{\mathcal{H}^{\prime }}$
in $\mathfrak{S}(\mathcal{H}^{\prime })$ and $\lambda=\log d'$.}

\textbf{Proof.} The first part of the corollary directly follows
from proposition 4.

Let $\Phi$ be a bistochastic channel. It is clear that
$\Omega(\Phi)=d^{\prime -1}\mathcal{I}_{\mathcal{H}^{\prime }}$
implies equality (\ref{c-cond-2}) with $\lambda=\log d'$.
Conversely, let equality (\ref{c-cond-2}) be true with some
$\lambda$ and $\bar{\rho}$ be the average state of an optimal
ensemble for the channel $\Phi$. Multiplying both sides of
equality (\ref{c-cond-2}) by $\bar{\rho}$ and taking the trace we
obtain $H(\Omega(\Phi))=\lambda$, and, hence $\lambda\leq\log d'$.

Due to
$\Phi(d^{-1}\mathcal{I}_{\mathcal{H}})=d^{\prime-1}\mathcal{I}_{\mathcal{H}'}$
we have
$$
\mathrm{Tr}\Phi(d^{-1}\mathcal{I}_{\mathcal{H}})\log\Phi(d^{-1}\mathcal{I}_{\mathcal{H}})=
-\log d'.
$$
Equality (\ref{c-cond-2}) implies
$$
-\mathrm{Tr}\Phi(d^{-1}\mathcal{I}_{\mathcal{H}})\log\Omega(\Phi)=
\mathrm{Tr}d^{-1}\mathcal{I}_{\mathcal{H}}\Phi^{*}(-\log\Omega(\Phi))=\lambda.
$$
Adding two last equalities we have
$$
H(\Phi(d^{-1}\mathcal{I}_{\mathcal{H}})\|\Omega(\Phi))=\lambda-\log
d'\leq 0.
$$
By the basic properties of the relative entropy
\cite{{Sch-West-2}} this inequality implies
$\Omega(\Phi)=\Phi(d^{-1}\mathcal{I}_{\mathcal{H}})=d^{\prime-1}\mathcal{I}_{\mathcal{H}'}$.
$\square$

The important class of channels for which condition (\ref{c-cond-2})
in corollary 1 is valid consists of channels covariant under the
action of irreducible representation of a symmetry group \cite{H-2}.

\section{Additivity properties}

Suppose we have two channels $\Phi
:\mathfrak{S}(\mathcal{H})\mapsto \mathfrak{S}(\mathcal{H}^{\prime
})$ and $\Psi :\mathfrak{S}(\mathcal{K})\mapsto
\mathfrak{S}(\mathcal{K}^{\prime })$. Denote by
$\Theta_{\mathcal{H}}$ and $\Theta_{\mathcal{K}}$ the mappings
$\omega\mapsto\omega^{\mathcal{H}}=\mathrm{Tr}_{\mathcal{K}}\omega$
and
$\omega\mapsto\omega^{\mathcal{K}}=\mathrm{Tr}_{\mathcal{H}}\omega$
from $\mathfrak{S}(\mathcal{H}\otimes\mathcal{K})$ onto
$\mathfrak{S}(\mathcal{H})$ and $\mathfrak{S}(\mathcal{K})$
correspondingly. For arbitrary convex subset $\mathcal{A}$ in
$\mathfrak{S}(\mathcal{H}\otimes\mathcal{K})$ the sets
$\Theta_{\mathcal{H}}(\mathcal{A})$ and
$\Theta_{\mathcal{K}}(\mathcal{A})$ are convex subsets of
$\mathfrak{S}(\mathcal{H})$ and of $\mathfrak{S}(\mathcal{K})$
correspondingly.

Consider the channel
$\Phi\otimes\Psi:\mathfrak{S}(\mathcal{H}\otimes\mathcal{K})\mapsto\mathfrak{S}(\mathcal{H}'\otimes\mathcal{K}')$.

Additivity of the  minimal output entropy for two channels $\Phi$
and $\Psi$ means
$$
H_{\mathrm{min}}(\Phi \otimes
\Psi)=H_{\mathrm{min}}(\Phi)+H_{\mathrm{min}}(\Psi).
$$
The following proposition provides the necessary and sufficient
condition for the additivity of the minimal output entropy in terms
of the sets $\mathcal{A}_{E}^{\Phi\otimes\Psi}$,
$\Theta_{\mathcal{H}}(\mathcal{A}_{E}^{\Phi\otimes\Psi})$,
$\Theta_{\mathcal{K}}(\mathcal{A}_{E}^{\Phi\otimes\Psi})$.

\textbf{Proposition 5.}\textit{ Additivity of the minimal output
entropy holds for the channels} $\Phi$ \textit{and} $\Psi$
\textit{if and only if one of the following equivalent conditions
holds}
\begin{itemize}
  \item \textit{there exists an unentangled extreme
   point of} $\mathcal{A}_{E}^{\Phi\otimes\Psi}$;
  \item
  \textit{$\mathcal{A}_{E}^{\Phi}\subseteq\Theta_{\mathcal{H}}(\mathcal{A}_{E}^{\Phi\otimes\Psi})\,$ and
  $\, \mathcal{A}_{E}^{\Psi}\subseteq\Theta_{\mathcal{K}}(\mathcal{A}_{E}^{\Phi\otimes\Psi})$.}

 \end{itemize}

\textit{Any pure state in
$\Theta_{\mathcal{H}}(\mathcal{A}_{E}^{\Phi\otimes\Psi})$ (in
$\Theta_{\mathcal{K}}(\mathcal{A}_{E}^{\Phi\otimes\Psi})$) lies in
$\mathcal{A}_{E}^{\Phi}$ (in $\mathcal{A}_{E}^{\Psi}$).
}\vspace{10pt}

\textbf{Proof.} This directly follows from the definition of the
sets $\mathcal{A}_{E}^{\Phi\otimes\Psi}$, $\mathcal{A}_{E}^{\Phi}$
and $\mathcal{A}_{E}^{\Psi}$. $\square$

Additivity of the Holevo capacity for  two channels $\Phi$ and
$\Psi$ means that
$$
\bar{C}(\Phi \otimes \Psi)=\bar{C}(\Phi)+\bar{C}(\Psi).
$$
The following proposition provides the necessary and sufficient
condition for the additivity of the Holevo capacity in terms of the
sets $\mathcal{A}_{C}^{\Phi\otimes\Psi}$,
$\Theta_{\mathcal{H}}(\mathcal{A}_{C}^{\Phi\otimes\Psi})$,
$\Theta_{\mathcal{K}}(\mathcal{A}_{C}^{\Phi\otimes\Psi})$.

\textbf{Proposition 6.} \textit{Additivity of the Holevo capacity
holds for the channels $\Phi$ and $\Psi$ if and only if
$\Omega(\Phi\otimes\Psi)=\Omega(\Phi)\otimes\Omega(\Psi)$ and one
of the following conditions holds:}

\begin{itemize}
  \item \textit{there exists an unentangled extreme point of
  $\mathcal{A}_{C}^{\Phi\otimes\Psi}$;}
  \item
  \textit{$\mathcal{A}_{C}^{\Phi}\subseteq\Theta_{\mathcal{H}}(\mathcal{A}_{C}^{\Phi\otimes\Psi})\,$ and
  $\, \mathcal{A}_{C}^{\Psi}\subseteq\Theta_{\mathcal{K}}(\mathcal{A}_{C}^{\Phi\otimes\Psi})$.}
\end{itemize}

\textit{If
$\Omega(\Phi\otimes\Psi)=\Omega(\Phi)\otimes\Omega(\Psi)$ then any
pure state in
$\Theta_{\mathcal{H}}(\mathcal{A}_{C}^{\Phi\otimes\Psi})$ (in
$\Theta_{\mathcal{K}}(\mathcal{A}_{C}^{\Phi\otimes\Psi})$) lies in
$\mathcal{A}_{C}^{\Phi}$ (in
$\mathcal{A}_{C}^{\Psi}$).}\vspace{10pt}

\textbf{Proof.} Additivity of the Holevo capacity for the channels
$\Phi$ and $\Psi$ obviously implies
$\Omega(\Phi\otimes\Psi)=\Omega(\Phi)\otimes\Omega(\Psi)$ and the
existence of product pure state in
$\mathcal{A}_{C}^{\Phi\otimes\Psi}$. To prove the implications
$\mathcal{A}_{C}^{\Phi}\subseteq\Theta_{\mathcal{H}}(\mathcal{A}_{C}^{\Phi\otimes\Psi})$
and
$\mathcal{A}_{C}^{\Psi}\subseteq\Theta_{\mathcal{K}}(\mathcal{A}_{C}^{\Phi\otimes\Psi})$
it is sufficient to show that
$\rho\otimes\sigma\in\mathcal{A}_{C}^{\Phi\otimes\Psi}$ for
arbitrary extreme points $\rho$ and $\sigma$ of
$\mathcal{A}_{C}^{\Phi}$ and of $\mathcal{A}_{C}^{\Psi}$
correspondingly. But this follows from proposition 3 and the
equality

$$
\begin{array}{c}
H(\Phi\otimes\Psi(\rho\otimes\sigma)\Vert
\Omega(\Phi\otimes\Psi))=H(\Phi(\rho)\otimes\Psi(\sigma)\Vert
\Omega(\Phi)\otimes\Omega(\Psi))\\\\=H(\Phi(\rho)\Vert
\Omega(\Phi))+H(\Psi(\sigma) \Vert\Omega(\Psi))=
\bar{C}(\Phi)+\bar{C}(\Psi)=\bar{C}(\Phi\otimes\Psi).
\end{array}
$$

To prove the converse statement note first that each of the
implications
$\mathcal{A}_{C}^{\Phi}\subseteq\Theta_{\mathcal{H}}(\mathcal{A}_{C}^{\Phi\otimes\Psi})$
and
$\mathcal{A}_{C}^{\Psi}\subseteq\Theta_{\mathcal{K}}(\mathcal{A}_{C}^{\Phi\otimes\Psi})$
implies existence of pure product state in
$\mathcal{A}_{C}^{\Phi\otimes\Psi}$.

Let $\Omega(\Phi\otimes\Psi)=\Omega(\Phi)\otimes\Omega(\Psi)$ and
$\rho\otimes\sigma$ be a pure state in
$\mathrm{Extr}\mathcal{A}_{C}^{\Phi\otimes\Psi}$. By proposition
3, we have
$$
\begin{array}{c}
\bar{C}(\Phi\otimes\Psi)=H(\Phi\otimes\Psi(\rho\otimes\sigma)\Vert
\Omega(\Phi\otimes\Psi))=H(\Phi(\rho) \otimes \Psi(\sigma) \Vert
\Omega(\Phi)\otimes\Omega(\Psi))\\\\=H(\Phi(\rho) \Vert
\Omega(\Phi))+H(\Psi(\sigma) \Vert\Omega(\Psi))\leq
\bar{C}(\Phi)+\bar{C}(\Psi ),
\end{array}
$$
where the last inequality follows from the maximal distance
property. Since the converse inequality is  obvious there is a
equality here, which implies
$H(\Phi(\rho)\Vert\Omega(\Phi))=\bar{C}(\Phi)$ and
$H(\Phi(\sigma)\Vert\Omega(\Phi))=\bar{C}(\Psi)$. By proposition 3
this means $\rho\in\mathcal{A}_{C}^{\Phi}$ and
$\sigma\in\mathcal{A}_{C}^{\Psi}$. This observation implies the
last statement of the proposition.$\square $

The condition
$\Omega(\Phi\otimes\Psi)=\Omega(\Phi)\otimes\Omega(\Psi)$ in the
above proposition holds for some special class of channels, in
particular, for irreducibly covariant channels, but it is
difficult to check in general. The following lemma provides the
sufficient condition for additivity  of the Holevo capacity for
the channels $\Phi$ and $\Psi$ in terms of extreme points of the
sets $\Theta_{\mathcal{H}}(\mathcal{A}_{C}^{\Phi\otimes\Psi})$ and
$\Theta_{\mathcal{K}}(\mathcal{A}_{C}^{\Phi\otimes\Psi})$.

\textbf{Lemma 2.} \textit{If there exists optimal ensemble for the
channel $\Phi\otimes\Psi$ with the product state average
$\tilde{\rho}\otimes\tilde{\sigma}$ and all the extreme points of
the sets $\Theta_{\mathcal{H}}(\mathcal{A}_{C}^{\Phi\otimes\Psi})$
and $\Theta_{\mathcal{K}}(\mathcal{A}_{C}^{\Phi\otimes\Psi})$ are
pure states then additivity  of the Holevo capacity holds for the
channels $\Phi$ and $\Psi$.}\footnote{It follows from the proof
that it is sufficient to require that the states $\tilde{\rho}$
and $\tilde{\sigma}$ can be represented as convex combinations of
pure states in
$\Theta_{\mathcal{H}}(\mathcal{A}_{C}^{\Phi\otimes\Psi})$ and
$\Theta_{\mathcal{K}}(\mathcal{A}_{C}^{\Phi\otimes\Psi})$
correspondingly.}

\textbf{Proof.} By definition 1
$\tilde{\rho}\otimes\tilde{\sigma}\in\mathcal{A}_{C}^{\Phi\otimes\Psi}$.
It implies
$\tilde{\rho}\in\Theta_{\mathcal{H}}(\mathcal{A}_{C}^{\Phi\otimes\Psi})$
and
$\tilde{\sigma}\in\Theta_{\mathcal{K}}(\mathcal{A}_{C}^{\Phi\otimes\Psi})$.
Hence $\tilde{\rho}=\sum_{i}\pi_{i}\rho_{i}$ and
$\tilde{\sigma}=\sum_{j}\mu_{j}\sigma_{j}$, where $\{\rho_{i}\}$ and
$\{\sigma_{j}\}$ are sets of extreme points(=pure states) of the
sets $\Theta_{\mathcal{H}}(\mathcal{A}_{C}^{\Phi\otimes\Psi})$ and
$\Theta_{\mathcal{K}}(\mathcal{A}_{C}^{\Phi\otimes\Psi})$.

To prove additivity of the Holevo capacity it is sufficient to
show that $\rho_{i}\otimes\sigma_{j}\in
\mathrm{Extr}\mathcal{A}_{C}^{\Phi\otimes\Psi}$ for all $i$ and
$j$. Indeed, if this holds, then, by proposition 3, the ensemble
$\{\pi_{i}\mu_{j}, \rho_{i}\otimes\sigma_{j}\}$ will be optimal
for the channel $\Phi\otimes\Psi$.

Let us fix some $\rho_{i}$ and $\sigma_{j}$. By the construction,
there exist pure states $\sigma_{i}\in \mathfrak{S}(\mathcal{K})$
and $\rho_{j}\in \mathfrak{S}(\mathcal{H})$ such that
$\rho_{i}\otimes\sigma_{i}\in
\mathrm{Extr}\mathcal{A}_{C}^{\Phi\otimes\Psi}$ and
$\rho_{j}\otimes\sigma_{j}\in
\mathrm{Extr}\mathcal{A}_{C}^{\Phi\otimes\Psi}$. By proposition 3
$$
H(\Phi(\rho_{i})\otimes\Psi(\sigma_{i})\|\Phi(\tilde{\rho})\otimes\Psi(\tilde{\sigma}))=
\bar{C}(\Phi\otimes\Psi)=
H(\Phi(\rho_{j})\otimes\Psi(\sigma_{j})\|\Phi(\tilde{\rho})\otimes\Psi(\tilde{\sigma}))
$$
Adding the above two equalities we obtain
$$
H(\Phi(\rho_{i})\otimes\Psi(\sigma_{j})\|\Phi(\tilde{\rho})\otimes\Psi(\tilde{\sigma}))+
H(\Phi(\rho_{j})\otimes\Psi(\sigma_{i})\|\Phi(\tilde{\rho})\otimes\Psi(\tilde{\sigma}))=
2\bar{C}(\Phi\otimes\Psi)
$$
By noting that each summing term in the left side does not exceed
$\bar{C}(\Phi\otimes\Psi)$ (the maximal distance property) we
conclude that
$$
H(\Phi(\rho_{i})\otimes\Psi(\sigma_{j})\|\Phi(\tilde{\rho})\otimes\Psi(\tilde{\sigma}))=
\bar{C}(\Phi\otimes\Psi)=
H(\Phi(\rho_{j})\otimes\Psi(\sigma_{i})\|\Phi(\tilde{\rho})\otimes\Psi(\tilde{\sigma})),
$$
which by proposition 3 implies $\rho_{i}\otimes\sigma_{j}\in
\mathrm{Extr}\mathcal{A}_{C}^{\Phi\otimes\Psi}$ and
$\rho_{j}\otimes\sigma_{i}\in
\mathrm{Extr}\mathcal{A}_{C}^{\Phi\otimes\Psi}$.$\square$

Let $\mathcal{A}$ and $\mathcal{B}$ be arbitrary subsets of
$\mathfrak{S}(\mathcal{H})$ and of $\mathfrak{S}(\mathcal{K})$
correspondingly. For the channel $\Phi\otimes \Psi$ it is natural
to consider the constraint defined by the requirements
$\bar{\omega}^{\mathcal{H}
}:=\mathrm{Tr}_{\mathcal{K}}\bar{\omega}\in \mathcal{A}$ and
$\bar{\omega}^{\mathcal{K}}:=\mathrm{Tr}_{\mathcal{H}
}\bar{\omega}\in \mathcal{B}$, where $\bar{\omega}$ is the average
state of an input ensemble $\{\mu _{i},\omega _{i}\}$. The subset
of $\mathfrak{S}(\mathcal{H}\otimes \mathcal{K})$ consisting of
all states $\omega$ such that $\omega^{\mathcal{H}}\in
\mathcal{A}$ and $\omega^{\mathcal{K}}\in \mathcal{B}$ will be
denoted $\mathcal{A}\otimes \mathcal{B}$. Additivity of the Holevo
capacity for the $\mathcal{A}$-constrained channel $\Phi$ and the
$\mathcal{B}$-constrained channel $\Psi$ means
$$
\bar{C}\left(\Phi \otimes \Psi ;\mathcal{A}\otimes
\mathcal{B}\right) =\bar{C}(\Phi ;\mathcal{A})+\bar{C}(\Psi
;\mathcal{B}).
$$

Validity of this additivity for arbitrary subsets $\mathcal{A}$
and $\mathcal{B}$ is substantially stronger than "unconstrained"
additivity and implies additivity of the minimal output entropy
\cite{Sh}. Hence, it is natural to call this property
\textit{strong additivity} of the Holevo capacity. In
\cite{H-Sh-1} it is shown that the class of channels for which
strong additivity of the Holevo capacity holds is nontrivial.

By proposition 5 additivity of the minimal output entropy for the
channels $\Phi$ and $\Psi$ implies
\begin{equation}\label{inclusion-1}
\mathcal{A}_{E}^{\Phi}\subseteq\Theta_{\mathcal{H}}(\mathcal{A}_{E}^{\Phi\otimes\Psi})\quad
\mathrm{and}
\quad\mathcal{A}_{E}^{\Psi}\subseteq\Theta_{\mathcal{K}}(\mathcal{A}_{E}^{\Phi\otimes\Psi}).
\end{equation}

By proposition 6 additivity of the Holevo capacity for the
channels $\Phi$ and $\Psi$ implies
\begin{equation}\label{inclusion-2}
\mathcal{A}_{C}^{\Phi}\subseteq\Theta_{\mathcal{H}}(\mathcal{A}_{C}^{\Phi\otimes\Psi})\quad
\mathrm{and}
\quad\mathcal{A}_{C}^{\Psi}\subseteq\Theta_{\mathcal{K}}(\mathcal{A}_{C}^{\Phi\otimes\Psi}).
\end{equation}

It follows that strong additivity of the Holevo capacity for the
channels $\Phi$ and $\Psi$ implies both the above inclusions
(\ref{inclusion-1}) and (\ref{inclusion-2}). But it turns out that
this property makes also possible to prove the converse inclusions
and hence to obtain the following \textit{ projective relations}
between optimal sets:
\begin{equation*}
\begin{array}{c}
\Theta_{\mathcal{H}}(\mathcal{A}_{C}^{\Phi\otimes\Psi})=\mathcal{A}_{C}^{\Phi},\quad
\Theta_{\mathcal{K}}(\mathcal{A}_{C}^{\Phi\otimes\Psi})=\mathcal{A}_{C}^{\Psi},\\\\
\Theta_{\mathcal{H}}(\mathcal{A}_{E}^{\Phi\otimes\Psi})=\mathcal{A}_{E}^{\Phi},\quad
\Theta_{\mathcal{K}}(\mathcal{A}_{E}^{\Phi\otimes\Psi})=\mathcal{A}_{E}^{\Psi}.
\end{array}
\end{equation*}

This relations are established in section 5 (proposition 8) on the
base of some structural results presented in the next section.

\section{On hereditary subsets of states}

In this section we consider subsets of states of a bipartite
system with the special properties, which seems to be of
independent interest.

\textbf{Definition 3.} \textit{The set
$\mathcal{A}\subseteq\mathfrak{S}(\mathcal{H}\otimes\mathcal{K})$
is called hereditary if
\[
\omega \in \mathcal{A}\;\Rightarrow
\;\omega^{\mathcal{H}}\otimes\omega^{\mathcal{K}}\in\mathcal{A}.
\]
The set
$\mathcal{A}\subseteq\mathfrak{S}(\mathcal{H}\otimes\mathcal{K})$
is called strong hereditary if
\[
\omega_{1},\omega_{2} \in \mathcal{A}\;\Rightarrow
\;\omega_{i}^{\mathcal{H}}\otimes \omega_{j}^{\mathcal{K}}\in
\mathcal{A},\;i,j=1,2.
\]}
A nontrivial example of strong hereditary set is provided by the
output set for arbitrary tensor product channel.

Hereditary properties of a subset $\mathcal{A}$ with pure extreme
points impose restrictions on the structure of the sets
$\Theta_{\mathcal{H}}(\mathcal{A})$ and
$\Theta_{\mathcal{K}}(\mathcal{A})$.

\textbf{Theorem.} \textit{If} $\mathcal{A}\subseteq
\mathfrak{S}(\mathcal{H}\otimes\mathcal{K})$ \textit{is a
hereditary convex set such that}
$\mathrm{Extr}\mathcal{A}\subseteq\mathrm{Extr}
\mathfrak{S}(\mathcal{H}\otimes\mathcal{K})$, \textit{then every
point of} $\mathrm{Extr}\Theta_{\mathcal{H}}(\mathcal{A})$
\textit{and of} $\mathrm{Extr}\Theta_{\mathcal{K}}(\mathcal{A})$
\textit{is a multiple of projector.}

\textit{If, in addition, $\mathcal{A}$ is a strong hereditary set
then all points of
$\mathrm{Extr}\Theta_{\mathcal{H}}(\mathcal{A})$ and of
$\mathrm{Extr}\Theta_{\mathcal{K}}(\mathcal{A})$ are multiple of
projector of the same rank. }

\textbf{Proof.} For arbitrary extreme point $\rho$ of the set
$\Theta_{\mathcal{H}}(\mathcal{A})$ let $
\Theta_{\mathcal{H}}^{-1}(\rho )$ be the subset of all states
$\omega $ in $\mathcal{A}$ such that
$\Theta_{\mathcal{H}}(\omega)=\rho$. The set
$\Theta_{\mathcal{K}}(\Theta_{\mathcal{H}}^{-1}(\rho))$ is convex
subset in $\Theta_{\mathcal{K}}(\mathcal{A})$. Let $\sigma$ be an
arbitrary extreme point of this subset.

By the above construction, there exists a state $\omega \in
\mathcal{A}$ such that $\omega^{\mathcal{H}}=\rho$ and
$\omega^{\mathcal{K}}=\sigma $. By the hereditary property of the
set $\mathcal{A}$ this means that $\rho\otimes\sigma\in
\mathcal{A}$ and hence
\begin{equation}\label{decomp}
\rho \otimes \sigma =\sum_{i}\pi_{i}\omega_{i},
\end{equation}
where $\{\omega_{i}\}$ is a set of $\mathcal{A}$ - extreme points
(which are pure states by the condition). We obtain
$\rho=\sum_{i}\pi_{i}\omega_{i}^{\mathcal{H}}$ and $\sigma
=\sum_{i}\pi_{i}\omega_{i}^{\mathcal{K}} $. By the extremality of
$\rho $ in $\Theta_{\mathcal{H}}(\mathcal{A})$ we obtain
$\omega_{i}^{\mathcal{H}}=\rho \;\forall i$. This means that each
$\omega_{i}$ lies in $\Theta_{\mathcal{H}}^{-1}(\rho)$. By the
extremality of $\sigma$ in
$\Theta_{\mathcal{K}}(\Theta_{\mathcal{H}}^{-1}(\rho))$ we obtain
$\omega_{i}^{\mathcal{K}}=\sigma \;\forall i$.

Applying lemma 3 below to decomposition (\ref{decomp}) we obtain the
statement of the first part of the theorem.

Let $\rho$ and $\sigma$ be arbitrary extreme points  of
$\Theta_{\mathcal{H}}(\mathcal{A})$ and of
$\Theta_{\mathcal{K}}(\mathcal{A})$ correspondingly. By the strong
hereditary property of the set $\mathcal{A}$ the state
$\rho\otimes\sigma$ lies in $\mathcal{A}$ and hence can be
represented as a convex combination $\sum_{k}\mu_{k}\omega_{k}$ of
extreme points of $\mathcal{A}$. By taking partial traces we obtain
$$
\rho=\sum_{k}\mu_{k}\Theta_{\mathcal{H}}(\omega_{k}),\quad
\sigma=\sum_{k}\mu_{k}\Theta_{\mathcal{K}}(\omega_{k}).
$$

Due to the extremality of $\rho$ and $\sigma$ in
$\Theta_{\mathcal{H}}(\mathcal{A})$  and in
$\Theta_{\mathcal{K}}(\mathcal{A})$ correspondingly we can deduce
that $\Theta_{\mathcal{H}}(\omega_{k})=\rho$ and
$\Theta_{\mathcal{K}}(\omega_{k})=\sigma$. Purity of the states
$\omega_{k}$ and the Schmidt decomposition implies equality of ranks
of the states $\rho$ and $\sigma$.

The above consideration with fixed state
$\rho\in\mathrm{Extr}\Theta_{\mathcal{H}}(\mathcal{A})$ and
arbitrary state
$\sigma\in\mathrm{Extr}\Theta_{\mathcal{K}}(\mathcal{A})$ implies
equality of rank of all states in
$\mathrm{Extr}\Theta_{\mathcal{K}}(\mathcal{A})$. Fixing arbitrary
state $\sigma\in\mathrm{Extr}\Theta_{\mathcal{K}}(\mathcal{A})$
and taking arbitrary state
$\rho\in\mathrm{Extr}\Theta_{\mathcal{H}}(\mathcal{A})$ we obtain
equality of ranks of all states in
$\mathrm{Extr}\Theta_{\mathcal{H}}(\mathcal{A})\cup\mathrm{Extr}\Theta_{\mathcal{K}}(\mathcal{A})$.$\square$

\textbf{Lemma 3.} \textit{A state} $\rho \otimes \sigma$ \textit{can
be represented as} $\,\sum_i\pi_{i}|\varphi _i\rangle \langle
\varphi _i|\,$\textit{, where} $\nobreak\mathrm{Tr}_{
\mathcal{K}}|\varphi _i\rangle \langle \varphi _i|=\rho $
\textit{and} $\mathrm{Tr}_{ \mathcal{H}}|\varphi _i\rangle \langle
\varphi _i|=\sigma $ \textit{if and only if} $ \rho =r^{-1} P$
\textit{and} $\sigma =r^{-1} Q$\textit{, where} $ P $ \textit{and} $
Q$ \textit{are projectors with} $\dim
 P(\mathcal{H})=\dim
 Q(\mathcal{K})=r$\textit{.}

\textbf{Proof.} Suppose the above decomposition takes place. It is
sufficient to prove that the state $\rho$ has no different positive
eigenvalues. Suppose $ \lambda _1$ and $\lambda _2$ are such
eigenvalues. Let $ P_1$ and $ Q_2$ be the corresponding spectral
projectors of the operators $\rho $ and $\sigma$.

By using the Schmidt decomposition for any vector $|\varphi
_i\rangle $ we have
\[
\left(  P_1\otimes  Q_2\right) |\varphi _i\rangle=\left(
 P_1\otimes
 Q_2\right)\sum_{j}\lambda_{j}|e^{i}_{j}\otimes
f^{i}_{j}\rangle=0,
\]
where $\{|e^{i}_{j}\rangle\}$ and $\{|f^{i}_{j}\rangle\}$  are some
orthonormal bases (for given $i$) of eigenvectors for $\rho$ and
$\sigma$ with the corresponding eigenvalues $\{\lambda_{j}\}$. Hence
$$
\begin{array}{c}
 0=\left(  P_1\otimes  Q_2\right)\sum_{i}\pi_{i}|\varphi_{i}\rangle\langle\varphi_{i}|=
\left(P_{1}\otimes Q_{2}\right) \left( \rho \otimes \sigma
\right)\\\\ = P_{1}\rho \otimes  Q_{2}\sigma
=\lambda_{1}\lambda_{2}\left(P_{1}\otimes Q_{2}\right),
\end{array}
$$
which is a contradiction.

It is sufficient to prove the converse statement of the lemma  in
the case $  P=\mathcal{I}_{\mathcal{H}}$ and $
Q=\mathcal{I}_{\mathcal{K}}$. The role of the set $\{|\varphi
_i\rangle \}$ in this case is played by orthonormal basis of
maximally entangled vectors in $\mathcal{H}\otimes
\mathcal{K}$.$\square $

\textbf{Examples.} Let $\mathcal{H}_{0}$ and $\mathcal{K}_{0}$ be
arbitrary subspaces of the spaces $\mathcal{H}$ and $\mathcal{K}$
correspondingly.

The set $\mathfrak{S}(\mathcal{H}_{0}\otimes\mathcal{K}_{0})$ of
all states in $\mathfrak{S}(\mathcal{H}\otimes\mathcal{K})$
supported by the subspace $\mathcal{H}_{0}\otimes\mathcal{K}_{0}$
is a strong hereditary subset of
$\mathfrak{S}(\mathcal{H}\otimes\mathcal{K})$ with pure extreme
points. In this case all the extreme points of the sets
$\Theta_{\mathcal{H}}(\mathfrak{S}(\mathcal{H}_{0}\otimes\mathcal{K}_{0}))=\mathfrak{S}(\mathcal{H}_{0})$
and
$\Theta_{\mathcal{K}}(\mathfrak{S}(\mathcal{H}_{0}\otimes\mathcal{K}_{0}))=\mathfrak{S}(\mathcal{K}_{0})$
are one dimensional projectors - pure states.

If the subspaces $\mathcal{H}_{0}$ and $\mathcal{K}_{0}$ have the
same dimension $r$ then the convex hull
$\mathfrak{M}(\mathcal{H}_{0}\otimes\mathcal{K}_{0})$ of all
maximally entangled states in
$\mathfrak{S}(\mathcal{H}_{0}\otimes\mathcal{K}_{0})$ is a strong
hereditary subset of $\mathfrak{S}(\mathcal{H}\otimes\mathcal{K})$
with pure extreme points. In this case
$\Theta_{\mathcal{H}}(\mathfrak{M}(\mathcal{H}_{0}\otimes\mathcal{K}_{0}))=\{r^{-1}
P\}$ and
$\Theta_{\mathcal{K}}(\mathfrak{M}(\mathcal{H}_{0}\otimes\mathcal{K}_{0}))=\{r^{-1}
Q\}$, where $ P$ and $ Q$ are the projectors on the subspaces
$\mathcal{H}_{0}$ and $\mathcal{K}_{0}$ correspondingly.

More sophisticated examples of hereditary sets with pure extreme
points can be constructed from the above simple examples as follows.
Let $\{\mathcal{H}_{i}\}$ and $\{\mathcal{K}_{j}\}$ are collections
of subspaces of the spaces $\mathcal{H}$ and $\mathcal{K}$
correspondingly. It is easy to see that the convex hull of the
collection of the sets
$\{\mathfrak{S}(\mathcal{H}_{i}\otimes\mathcal{K}_{j})\}$ is a
strong hereditary subset of
$\mathfrak{S}(\mathcal{H}\otimes\mathcal{K})$ with pure extreme
points. If all the subspaces in the above collections have the same
dimension then the convex hull of the collection of the sets
$\{\mathfrak{M}(\mathcal{H}_{i}\otimes\mathcal{K}_{j})\}$ is a
strong hereditary subset of
$\mathfrak{S}(\mathcal{H}\otimes\mathcal{K})$ with pure extreme
points.$\square$

The above theorem implies that inside any convex hereditary set
$\mathcal{A}\subseteq\mathfrak{S}(\mathcal{H}\otimes\mathcal{K})$
with pure extreme points there exist "sufficiently many" pure states
whose partial traces are multiple of projectors. By the Schmidt
decomposition such pure states are generated by vectors of the
following form
\begin{equation}\label{u-e-v}
|\varphi\rangle=\frac{1}{\sqrt{r}}\sum_{i=1}^{r}|e_{i}\otimes
f_{i}\rangle,
\end{equation}
where $\{|e_{i}\rangle\}$ and $\{|f_{i}\rangle\}$ are some
orthonormal systems of vectors in $\mathcal{H}$ and $\mathcal{K}$
correspondingly. In the case $r=\dim\mathcal{H}=\dim\mathcal{K}$
the vector of the form (\ref{u-e-v}) is referred to as maximally
entangled. Generalizing this terminology we introduce the
following definition.

\textbf{Definition 4.} \textit{The vector of the form
(\ref{u-e-v}) is called uniformly entangled vector of rank $r$. }

In the following section we will see that under the particular
conditions the optimal sets (introduced in section 2) for a tensor
product channel have the hereditary and the strong hereditary
properties.

\section{On the structure of the sets $\mathcal{A}_{C}^{\Phi\otimes\Psi}$ and $\mathcal{A}_{E}^{\Phi\otimes\Psi}$}

In this section we explore the properties of the sets
$\mathcal{A}_{C}^{\Phi\otimes\Psi}$ and
$\mathcal{A}_{E}^{\Phi\otimes\Psi}$ for tensor product channel
$\Phi\otimes\Psi$ under the two following assumptions:

\begin{enumerate} [A)]

  \item For arbitrary constraint sets $\mathcal{A}$ and
  $\mathcal{B}$ there exists optimal ensemble for the
  $\mathcal{A}\otimes\mathcal{B}$-constrained channel $\Phi\otimes\Psi$ with the
  product state average;

  \item Strong additivity of the Holevo capacity holds for the
   channels $\Phi$ and $\Psi$.

\end{enumerate}

Note that the assumption B can be reformulated in the manner
similar to the assumption A:

\begin{enumerate} [B')]
  \item For arbitrary
constraint sets $\mathcal{A}$ and $\mathcal{B}$ there exists optimal
ensemble for the $\mathcal{A}\otimes\mathcal{B}$-constrained channel
$\Phi\otimes\Psi$ with the product state average consisting of
product states.
\end{enumerate}

\subsection{Assumption A}

The above observation shows that the assumption A can be
considered as the weak form of the assumption B. It is easy to see
that the assumption A is equivalent to the validity of the
following inequality
\begin{equation}\label{main}
\chi_{\Phi\otimes\Psi}(\omega^{\mathcal{H}}\otimes\omega^{\mathcal{K}})\geq
\chi_{\Phi\otimes\Psi}(\omega)
\end{equation}
for all states $\omega\in
\mathfrak{S}(\mathcal{H}\otimes\mathcal{K})$.

\textbf{Proposition 7.} \textit{If the assumption A holds for the
channels $\Phi$ and $\Psi$ then the sets
$\mathcal{A}_{C}^{\Phi\otimes\Psi}$ and
$\mathcal{A}_{E}^{\Phi\otimes\Psi}$ are hereditary subsets of
$\mathfrak{S}(\mathcal{H}\otimes\mathcal{K})$ with pure extreme
points  and for arbitrary state $\omega$ of these sets the
following property holds
$$
\hat{H}_{\Phi\otimes\Psi}(\omega)=
\hat{H}_{\Phi\otimes\Psi}(\omega^{\mathcal{H}}\otimes\omega^{\mathcal{K}}).
$$}

\textbf{Proof.} Note first that validity of inequality
(\ref{main}) implies validity of the following inequality
\begin{equation}\label{main-1}
\chi_{\Phi\otimes\Psi}(\omega^{\mathcal{H}}\otimes\omega^{\mathcal{K}})-\chi
_{\Phi\otimes\Psi}(\omega )\geq H(\Phi\otimes\Psi(\omega)
||\Phi(\omega^{\mathcal{H}})\otimes\Psi(\omega^{\mathcal{K}})).
\end{equation}
Indeed, this follows from corollary 1 in \cite{H-Sh-1} since
validity of inequality (\ref{main}) for a state $\omega$  means
that $\omega^{\mathcal{H}}\otimes\omega^{\mathcal{K}}$ is the
average state of some optimal ensemble for the channel
$\Phi\otimes\Psi$ with the constraint defined by the convex set
$\{\omega^{\mathcal{H}}\}\otimes\{\omega^{\mathcal{K}}\}$.

Let us prove the statement of the proposition for
$\mathcal{A}_{C}^{\Phi\otimes\Psi}$. By proposition 3 all extreme
points of this set are pure states. By the assumption A there exists
optimal ensemble for the unconstrained channel $\Phi\otimes\Psi$
with the average state $\tilde{\rho}\otimes \tilde{\sigma}$. Let
$\omega$ be an arbitrary state in
$\mathcal{A}_{C}^{\Phi\otimes\Psi}$. This means that
\begin{equation}
\chi _{\Phi\otimes\Psi}(\omega )=\bar{C}(\Phi \otimes \Psi
)-H(\Phi\otimes\Psi(\omega) ||\Phi(\tilde{\rho})\otimes
\Psi(\tilde{\sigma})). \label{one}
\end{equation}

By inequalities (\ref{chi-ineq}) and (\ref{main-1}) we have
\begin{equation}\label{two}
\begin{array}{ccl}
\chi_{\Phi\otimes\Psi}(\omega) & \leq & \chi
_{\Phi\otimes\Psi}(\omega^{\mathcal{H}}\otimes\omega^{\mathcal{K}})-H(\Phi\otimes\Psi(\omega)
||\Phi(\omega^{\mathcal{H}})\otimes\Psi(\omega^{\mathcal{K}}))\\\\&\leq
&\bar{C}(\Phi\otimes \Psi )-H(\Phi(\omega^{\mathcal{H}})\otimes
\Psi(\omega^{\mathcal{K}})\Vert \Phi(\tilde{\rho})\otimes
\Psi(\tilde{\sigma}))\\\\&-&H(\Phi\otimes\Psi(\omega)
||\Phi(\omega^{\mathcal{H}})\otimes\Psi(\omega^{\mathcal{K}}))\\\\&=&\bar{C}(\Phi
\otimes \Psi )-H(\Phi\otimes\Psi(\omega) \Vert
\Phi(\tilde{\rho})\otimes \Psi(\tilde{\sigma})).
\end{array}
\end{equation}

Comparing (\ref{one}) and (\ref{two}) we obtain that all
inequalities in (\ref{two}) are in fact equalities. By the
definition the second equality in (\ref{two}) means that
$\omega^{\mathcal{H}}\otimes\omega^{\mathcal{K}}\in\mathcal{A}_{C}^{\Phi\otimes\Psi}$
while the first equality in (\ref{two}) can be rewritten as
$$
\chi_{\Phi\otimes\Psi}(\omega)-H(\Phi\otimes\Psi(\omega)) = \chi
_{\Phi\otimes\Psi}(\omega^{\mathcal{H}}\otimes\omega^{\mathcal{K}})
-H(\Phi(\omega^{\mathcal{H}})\otimes\Psi(\omega^{\mathcal{K}})),
$$
which implies the second statement of the proposition in the case
of $\mathcal{A}_{C}^{\Phi\otimes\Psi}$.

Let us prove the statement of the proposition for
$\mathcal{A}_{E}^{\Phi\otimes\Psi}$. By the definition all extreme
points of the this set are pure states.

Suppose inequality (\ref{main}) and, hence, inequality
(\ref{main-1}) hold for arbitrary $\omega$ in
$\mathcal{A}_{E}^{\Phi\otimes\Psi}$. The last inequality and
representation (\ref{chi-rep}) mean
$$
\begin{array}{c}
\widehat{H}_{\Phi\otimes\Psi}(\omega)\geq
\widehat{H}_{\Phi\otimes\Psi}(\omega^{\mathcal{H}}\otimes\omega^{\mathcal{K}}).
\end{array}
$$
By the definition of the set $\mathcal{A}_{E}^{\Phi\otimes\Psi}$
this implies equality
$\hat{H}_{\Phi\otimes\Psi}(\omega^{\mathcal{H}}\otimes\omega^{\mathcal{K}})
=\hat{H}_{\Phi\otimes\Psi}(\omega)=H_{\mathrm{min}}(\Phi\otimes\Psi)$
and decomposition of the state
$\omega^{\mathcal{H}}\otimes\omega^{\mathcal{K}}$ as a convex
combination of extreme points of
$\mathcal{A}_{E}^{\Phi\otimes\Psi}$. By definition this means that
$\omega^{\mathcal{H}}\otimes\omega^{\mathcal{K}}\in\mathcal{A}_{E}^{\Phi\otimes\Psi}$.$\square$

\textbf{Corollary 2.} \textit{If the assumption A holds for the
channels} $\Phi$ \textit{and }$\Psi$ \textit{then there exists an
 uniformly entangled state} $\omega$
\textit{optimal for the channel $\Phi\otimes\Psi$ and such that
any other optimal state} $\omega'$ \textit{with}
$$
\mathrm{supp}\mathrm{Tr}_{\mathcal{H}}\omega'\subseteq
\mathrm{supp}\mathrm{Tr}_{\mathcal{H}}\omega,\quad
\mathrm{supp}\mathrm{Tr}_{\mathcal{K}}\omega'\subseteq
\mathrm{supp}\mathrm{Tr}_{\mathcal{K}}\omega.
$$
\textit{is also uniformly entangled with the same rank and}
$$
\mathrm{Tr}_{\mathcal{H}}\omega'=\mathrm{Tr}_{\mathcal{H}}\omega,
\quad\mathrm{Tr}_{\mathcal{K}}\omega'=\mathrm{Tr}_{\mathcal{K}}\omega.
$$

\textbf{Proof.} Note that in the case of the channels $\Phi$ and
$\Psi$, for which the additivity of the minimal entropy holds, the
statement of the proposition is obvious, because  any nonentangled
optimal pure state for $\Phi\otimes\Psi$ is uniformly entangled
state of rank 1.

By proposition 7 $\,\mathcal{A}_{E}^{\Phi\otimes\Psi}$ is a
hereditary subset of $\mathfrak{S}(\mathcal{H}\otimes\mathcal{K})$
with pure extreme points. The theorem in section 4 implies  that all
extreme points of the sets
$\Theta_{\mathcal{H}}(\mathcal{A}_{E}^{\Phi\otimes\Psi})$ and
$\Theta_{\mathcal{K}}(\mathcal{A}_{E}^{\Phi\otimes\Psi})$ are
multiple of projectors. By proposition 5 nonadditivity of the
minimal output entropy implies that all these projectors can not be
one-dimensional. Let $r^{-1} P$ be an extreme point of
$\Theta_{\mathcal{H}}(\mathcal{A}_{E}^{\Phi\otimes\Psi})$ of the
minimal rank $r\geq 2$. This means that there exists a pure state
$\omega\in\mathrm{Extr}\mathcal{A}_{E}^{\Phi\otimes\Psi}$ such that
$\omega^{\mathcal{H}}=r^{-1} P$. It follows that
$Q=r\omega^{\mathcal{K}}$ is a projector of rank $r$. Let
$\mathcal{H}_{0}=P(\mathcal{H})$ and
$\mathcal{K}_{0}=Q(\mathcal{K})$. By the choice of $P$ for all
$\omega$ in $\mathrm{Extr}\mathcal{A}_{E}^{\Phi\otimes\Psi}$ with
$\mathrm{supp}\omega^{\mathcal{H}}\subseteq\mathcal{H}_{0}$ it holds
$\omega^{\mathcal{H}}=r^{-1} P$. Indeed, converse implies existence
of the projector type extreme point in
$\Theta_{\mathcal{H}}(\mathcal{A}_{E}^{\Phi\otimes\Psi})$ with rank
$<r$. $\square$

\textbf{Corollary 3.} \textit{If the assumption A holds for the
channels $\Phi$ and $\Psi$ while additivity of the minimal output
entropy does not hold then there exist subchannels $\Phi_{0}$ and
$\Psi_{0}$ of the channels $\Phi$ and $\Psi$ corresponding to the
nontrivial subspaces $\mathcal{H}_{0}$ and $\mathcal{K}_{0}$ with
the following properties\footnote{A subchannel corresponding to a
particular subspace is the restriction of the initial channel to
the set of states supported by this subspace \cite{Sh-2}.}:}
\begin{itemize}
  \item
  $H_{\mathrm{min}}(\Phi\otimes\Psi)=H_{\mathrm{min}}(\Phi_{0}\otimes\Psi_{0})$;
  \item \textit{any optimal state for the channel $\Phi_{0}\otimes\Psi_{0}$ is maximally
        entangled;}
  \item \textit{the chaotic state in $\mathfrak{S}(\mathcal{H}_{0}\otimes\mathcal{K}_{0})$
        can be represented as a convex combination of  maximally
        entangled states optimal for the channel $\Phi_{0}\otimes\Psi_{0}$.}
\end{itemize}

\textit{If, in addition, $\Phi$ and $\Psi$ are qubit channels than
$\Phi_{0}=\Phi$ and $\Psi_{0}=\Psi$.}\footnote{Additivity of the
minimal output entropy for unital qubit channels is proved by King
\cite{King}, but it validity for general qubit channel is an open
problem.} \vspace{5pt}

\textbf{Proof.} Let $\omega$ be a uniformly entangled optimal state,
existing by corollary 2. Since additivity of the minimal output
entropy does not hold for the channels $\Phi$ and $\Psi$ the rank of
this state is greater than 1. Let
$\mathcal{H}_{0}=\mathrm{supp}\mathrm{Tr}_{\mathcal{K}}\omega$ and
$\mathcal{K}_{0}=\mathrm{supp}\mathrm{Tr}_{\mathcal{H}}\omega$.
Reducing the channels $\Phi$ and $\Psi$ to
$\mathfrak{S}(\mathcal{H}_{0})$ and to
$\mathfrak{S}(\mathcal{K}_{0})$ correspondingly we obtain desired
subchannels $\Phi_{0}$ and $\Psi_{0}$. By the construction
$\omega^{\mathcal{H}_{0}}\otimes\omega^{\mathcal{K}_{0}}$ is the
chaotic state in
$\mathfrak{S}(\mathcal{H}_{0}\otimes\mathcal{K}_{0})$ and lies in
$\mathcal{A}_{E}^{\Phi\otimes\Psi}$ due to the hereditary property
of $\mathcal{A}_{E}^{\Phi\otimes\Psi}$.$\square$

\textbf{Corollary 4.} \textit{If the assumption A holds for the
channels} $\Phi$ \textit{and} $\Psi$. \textit{Then nonadditivity
of the Holevo capacity for these channels implies existence of
projectors} $ P\in\mathfrak{B}(\mathcal{H})$ \textit{and} $
Q\in\mathfrak{B}(\mathcal{K})$ \textit{of the same rank} $r\geq 2$
\textit{with the following properties:}

\begin{itemize}
  \item \textit{the state} $r^{-2} P\otimes Q$
   \textit{can be represented as convex combination of uniformly entangled states
   of rank} $r$\textit{,\footnote{This implies that each of these states has partial traces
   $r^{-1} P$ and $r^{-1} Q$.} having the same
   output entropy};
  \item \textit{this decomposition of the state} $r^{-2} P\otimes Q$ \textit{is
   optimal in the sense of the definition of the function}
   $\widehat{H}_{\Phi\otimes\Psi}$.
\end{itemize}

\textbf{Proof.} By lemma 2, proposition 7 and the theorem in section
4 nonadditivity of the Holevo capacity implies existence in
$\mathrm{Extr}\Theta_{\mathcal{H}}(\mathcal{A}_{C}^{\Phi\otimes\Psi})$
or in
$\mathrm{Extr}\Theta_{\mathcal{K}}(\mathcal{A}_{C}^{\Phi\otimes\Psi})$
of a state proportional to $r$-dimensional projector with $r\geq 2$.
Let $r^{-1}
P\in\mathrm{Extr}\Theta_{\mathcal{H}}(\mathcal{A}_{C}^{\Phi\otimes\Psi})$
be such a state. Hence there exists a pure state
$\omega\in\mathrm{Extr}\mathcal{A}_{C}^{\Phi\otimes\Psi}$ such that
$\omega^{\mathcal{H}}=r^{-1} P$. It follows that
$Q=r\omega^{\mathcal{K}}$ is a projector of rank $r$. Let
$\mathcal{H}_{0}=P(\mathcal{H})$ and
$\mathcal{K}_{0}=Q(\mathcal{K})$.

By hereditary property of $\mathcal{A}_{C}^{\Phi\otimes\Psi}$
(proposition 7) the state $r^{-2} P\otimes Q$ lies in
$\mathcal{A}_{C}^{\Phi\otimes\Psi}$ and hence can be represented
as convex combination of extreme points of
$\mathcal{A}_{C}^{\Phi\otimes\Psi}$. The partial traces of any
element of this decomposition must be equal to $r^{-1} P$ and
$r^{-1} Q$. Indeed, denote by prime all elements for which it is
not true. Then we have
$$
r^{-2} P\otimes
Q=\sum_{i}\pi_{i}\omega_{i}+\sum_{i}\pi'_{i}\omega'_{i}.
$$
Taking partial trace over space $\mathcal{K}$ we obtain
$$
r^{-1} P=\sum_{i}\pi_{i}r^{-1}
P+\sum_{i}\pi'_{i}\Theta_{\mathcal{H}}(\omega'_{i}).
$$
and hence
$$
r^{-1}
P=\left(1-\sum_{i}\pi_{i}\right)^{-1}\sum_{i}\pi'_{i}\Theta_{\mathcal{H}}(\omega'_{i})
=\left(\sum_{i}\pi'_{i}\right)^{-1}\sum_{i}\pi'_{i}\Theta_{\mathcal{H}}(\omega'_{i}).
$$
The last equality is a representation of $r^{-1}
P\in\mathrm{Extr}\Theta_{\mathcal{H}}(\mathcal{A}_{C}^{\Phi\otimes\Psi})$
as convex combination of states in
$\Theta_{\mathcal{H}}(\mathcal{A}_{C}^{\Phi\otimes\Psi})$. Hence
$\Theta_{\mathcal{H}}(\omega'_{i})=r^{-1} P$. Dimensionality
arguments and the Schmidt decomposition show that
$\Theta_{\mathcal{K}}(\omega'_{i})=r^{-1} Q$. Thus we obtain a
contradiction to the definition of the states $\omega'_{i}$. Hence
we have representation of the state $r^{-2} P\otimes
Q\in\mathcal{A}_{C}^{\Phi\otimes\Psi}$ as a convex combination
$\sum_{i}\pi_{i}\omega_{i}$ of states $\omega_{i}$ such that
$\omega_{i}^{\mathcal{H}}=r^{-1} P$ and
$\omega_{i}^{\mathcal{K}}=r^{-1} Q$. By proposition 7 this implies
$$
H_{\Phi\otimes\Psi}(\omega_{i})=\hat{H}_{\Phi\otimes\Psi}(\omega_{i})=
\hat{H}_{\Phi\otimes\Psi}(r^{-2} P\otimes Q)\quad \forall i,
$$
which completes the proof of the proposition. $\square$

\textbf{Corollary 5.} \textit{If for qubit channels} $\Phi$
\textit{and} $\Psi$ \textit{the assumption A holds while
additivity of the Holevo capacity does not hold then the chaotic
state} $\frac{1}{4}\mathcal{I}_{\mathcal{H}\otimes\mathcal{K}}$
\textit{in} $\mathfrak{S}(\mathcal{H}\otimes\mathcal{K})$
\textit{can be represented as a convex combination of maximally
entangled states with the same output entropy and this
decomposition of the chaotic state is optimal in the sense of the
definition of the function} $\widehat{H}_{\Phi\otimes\Psi}$.

\textbf{Proof.} In the case of two qubit channels $\Phi$ and
$\Psi$ every state of the form $d^{-2} P\otimes Q$ is either pure
or chaotic.$\square$

Let channels $\Phi $ and $\Psi $ be covariant under the
irreducible actions of unitary groups $\{U_{t}\}$ and $\{V_{s}\}$
correspondingly. Then the channel $\Phi \otimes \Psi $ is
covariant under irreducible action of the group $\{U_{t}\otimes
V_{s}\}$. In this case additivity of the minimal output entropy is
equivalent to additivity of the Holevo capacity. In \cite{H-2} it
was shown that any optimal ensemble for the channel
$\Phi\otimes\Psi$ has chaotic average and can be constructed from
one orbit of the group $\{U_{t}\otimes V_{s}\}$, containing a
state with minimal output entropy. This and corollary 2 imply the
following result.

\textbf{Corollary 6.} \textit{If the assumption A holds for
irreducibly covariant channels} $\Phi$ \textit{and} $\Psi$
\textit{then there exists an optimal ensemble for the channel
$\Phi\otimes\Psi$, consisting of uniformly entangled states of the
same rank.}

\textbf{Proof.} It is sufficient to note that the rank of
uniformly entangled states does not change under the action of the
group $\{U_{t}\otimes V_{s}\} $. $\square$

\subsection{Assumption B}

It is easy to see that the assumption B is equivalent to the
validity of the following inequality
\begin{equation}\label{main-B}
\chi_{\Phi\otimes\Psi}(\omega)\leq\chi_{\Phi}(\omega^{\mathcal{H}})+\chi_{\Psi}(\omega^{\mathcal{K}})
\end{equation}
for all states $\omega\in
\mathfrak{S}(\mathcal{H}\otimes\mathcal{K})$. By combining the
results in \cite{H-Sh-1} and \cite{P} it is possible to show that
this subadditivity property of the $\chi$-function is equivalent
to validity of the equalities
\begin{equation}\label{main-B-1}
\chi_{\Phi\otimes\Psi}(\rho\otimes\sigma)=\chi_{\Phi}(\rho)+\chi_{\Psi}(\sigma)
\end{equation}
and
\begin{equation}\label{main-B-2}
\hat{H}_{\Phi \otimes \Psi }(\rho\otimes\sigma)= \hat{H}_\Phi
(\rho)+ \hat{ H}_\Psi (\sigma)
\end{equation}
for arbitrary states $\rho\in \mathfrak{S}(\mathcal{H})$ and
$\sigma\in \mathfrak{S}(\mathcal{K})$.

\textbf{Proposition 8.} \textit{If the assumption B holds for the
channels $\Phi$ and $\Psi$ then the sets
$\mathcal{A}_{C}^{\Phi\otimes\Psi}$ and
$\mathcal{A}_{E}^{\Phi\otimes\Psi}$ are strong hereditary subsets of
$\mathfrak{S}(\mathcal{H}\otimes\mathcal{K})$ with pure extreme
points and the following "projective" relations take place
$$
\begin{array}{c}
\Theta_{\mathcal{H}}(\mathcal{A}_{C}^{\Phi\otimes\Psi})=\mathcal{A}_{C}^{\Phi},\quad
\Theta_{\mathcal{K}}(\mathcal{A}_{C}^{\Phi\otimes\Psi})=\mathcal{A}_{C}^{\Psi},\\\\
\Theta_{\mathcal{H}}(\mathcal{A}_{E}^{\Phi\otimes\Psi})=\mathcal{A}_{E}^{\Phi},\quad
\Theta_{\mathcal{K}}(\mathcal{A}_{E}^{\Phi\otimes\Psi})=\mathcal{A}_{E}^{\Psi}.
\end{array}
$$}

\textbf{Proof.} By proposition 7
$\,\mathcal{A}_{C}^{\Phi\otimes\Psi}$ and
$\mathcal{A}_{E}^{\Phi\otimes\Psi}$ are hereditary subsets of
$\mathfrak{S}(\mathcal{H}\otimes\mathcal{K})$ with pure extreme
points.

The assumption B implies
$\Omega(\Phi\otimes\Psi)=\Omega(\Phi)\otimes\Omega(\Psi)$. Let
$\omega_{1}$ and $\omega_{2}$ be arbitrary states in
$\mathcal{A}_{C}^{\Phi\otimes\Psi}$. The hereditary property implies
$\omega^{\mathcal{H}}_{i}\otimes\omega^{\mathcal{K}}_{i}\in\mathcal{A}_{C}^{\Phi\otimes\Psi},\quad
i=1,2$. By the definition of the set
$\mathcal{A}_{C}^{\Phi\otimes\Psi}$ it means
$$
\chi_{\Phi\otimes\Psi}(\omega^{\mathcal{H}}_{i}\otimes\omega^{\mathcal{K}}_{i})+
H(\Phi(\omega^{\mathcal{H}}_{i})\otimes\Psi(\omega^{\mathcal{K}}_{i})\|
\Omega(\Phi)\otimes\Omega(\Psi))= \bar{C}(\Phi\otimes\Psi) \quad
i=1,2.
$$

Equality (\ref{main-B-1}) implies
$$
\chi_{\Phi\otimes\Psi}(\omega^{\mathcal{H}}_{1}\otimes\omega^{\mathcal{K}}_{1})+
\chi_{\Phi\otimes\Psi}(\omega^{\mathcal{H}}_{2}\otimes\omega^{\mathcal{K}}_{2})=
\chi_{\Phi\otimes\Psi}(\omega^{\mathcal{H}}_{1}\otimes\omega^{\mathcal{K}}_{2})+
\chi_{\Phi\otimes\Psi}(\omega^{\mathcal{H}}_{2}\otimes\omega^{\mathcal{K}}_{1})
$$

Using this and adding the above two equalities we obtain
$$
\begin{array}{c}
\left[\chi_{\Phi\otimes\Psi}(\omega^{\mathcal{H}}_{1}\otimes\omega^{\mathcal{K}}_{2})+
H(\Phi(\omega^{\mathcal{H}}_{1})\otimes\Psi(\omega^{\mathcal{K}}_{2})\|
\Omega(\Phi)\otimes\Omega(\Psi))\right]\\\\+
\left[\chi_{\Phi\otimes\Psi}(\omega^{\mathcal{H}}_{2}\otimes\omega^{\mathcal{K}}_{1})+
H(\Phi(\omega^{\mathcal{H}}_{2})\otimes\Psi(\omega^{\mathcal{K}}_{1})\|
\Omega(\Phi)\otimes\Omega(\Psi))\right]=2\bar{C}(\Phi\otimes\Psi)
\end{array}
$$
Noting that each  term  in the square brackets does not exceed
$\bar{C}(\Phi\otimes\Psi)$ (corollary 1 in \cite{H-Sh-1}) we
conclude that
$$
\begin{array}{c}
\chi_{\Phi\otimes\Psi}(\omega^{\mathcal{H}}_{1}\otimes\omega^{\mathcal{K}}_{2})+
H(\Phi(\omega^{\mathcal{H}}_{1})\otimes\Psi(\omega^{\mathcal{K}}_{2})\|
\Omega(\Phi)\otimes\Omega(\Psi))=\bar{C}(\Phi\otimes\Psi),\\\\
\chi_{\Phi\otimes\Psi}(\omega^{\mathcal{H}}_{2}\otimes\omega^{\mathcal{K}}_{1})+
H(\Phi(\omega^{\mathcal{H}}_{2})\otimes\Psi(\omega^{\mathcal{K}}_{1})\|
\Omega(\Phi)\otimes\Omega(\Psi))= \bar{C}(\Phi\otimes\Psi).
\end{array}
$$

By definition this means that
$\omega^{\mathcal{H}}_{1}\otimes\omega^{\mathcal{K}}_{2}\in\mathcal{A}_{C}^{\Phi\otimes\Psi}$
and
$\omega^{\mathcal{H}}_{2}\otimes\omega^{\mathcal{K}}_{1}\in\mathcal{A}_{C}^{\Phi\otimes\Psi}$.
The strong hereditary property of the set
$\mathcal{A}_{C}^{\Phi\otimes\Psi}$ has proved.

Let $\omega_{1}$ and $\omega_{2}$ be arbitrary states in
$\mathcal{A}_{E}^{\Phi\otimes\Psi}$. Hereditary property implies
that
$\omega^{\mathcal{H}}_{i}\otimes\omega^{\mathcal{K}}_{i}\in\mathcal{A}_{E}^{\Phi\otimes\Psi},\quad
i=1,2$. By definition of the set
$\mathcal{A}_{E}^{\Phi\otimes\Psi}$ it means
$$
\hat{H}_{\Phi\otimes\Psi}(\omega^{\mathcal{H}}_{i}\otimes\omega^{\mathcal{K}}_{i})=H_{\mathrm{min}}(\Phi\otimes\Psi),
\quad i=1,2
$$
Due to equality (\ref{main-B-2}) we have
$$
\hat{H}_{\Phi\otimes\Psi}(\omega^{\mathcal{H}}_{1}\otimes\omega^{\mathcal{K}}_{1})+
\hat{H}_{\Phi\otimes\Psi}(\omega^{\mathcal{H}}_{2}\otimes\omega^{\mathcal{K}}_{2})=
\hat{H}_{\Phi\otimes\Psi}(\omega^{\mathcal{H}}_{1}\otimes\omega^{\mathcal{K}}_{2})+
\hat{H}_{\Phi\otimes\Psi}(\omega^{\mathcal{H}}_{2}\otimes\omega^{\mathcal{K}}_{1}).
$$
Using this and adding the above two equalities we obtain
$$
\hat{H}_{\Phi\otimes\Psi}(\omega^{\mathcal{H}}_{1}\otimes\omega^{\mathcal{K}}_{2})+
\hat{H}_{\Phi\otimes\Psi}(\omega^{\mathcal{H}}_{2}\otimes\omega^{\mathcal{K}}_{1})=2H_{\mathrm{min}}(\Phi\otimes\Psi),
$$
Noting that each  term in the left side is not less than
$H_{\mathrm{min}}$ we conclude that
$$
\begin{array}{c}
\hat{H}_{\Phi\otimes\Psi}(\omega^{\mathcal{H}}_{1}\otimes\omega^{\mathcal{K}}_{2})=H_{\mathrm{min}},\quad
\hat{H}_{\Phi\otimes\Psi}(\omega^{\mathcal{H}}_{2}\otimes\omega^{\mathcal{K}}_{1})=H_{\mathrm{min}}(\Phi\otimes\Psi).
\end{array}
$$
By definition this means that
$\omega^{\mathcal{H}}_{1}\otimes\omega^{\mathcal{K}}_{2}\in\mathcal{A}_{E}^{\Phi\otimes\Psi}$
and
$\omega^{\mathcal{H}}_{2}\otimes\omega^{\mathcal{K}}_{1}\in\mathcal{A}_{E}^{\Phi\otimes\Psi}$.
The strong hereditary property of the set
$\mathcal{A}_{E}^{\Phi\otimes\Psi}$ has also proved.

To prove the projective relations note that the assumption B implies
additivity of the minimal output entropy and additivity of the
Holevo capacity \cite{H-Sh-1}. This and proposition 5 and 6 imply
$"\supseteq"$ in the above projective relations and existence of at
least one nonentangled pure extreme point of the sets
$\Theta_{\mathcal{H}}(\mathcal{A}_{E}^{\Phi\otimes\Psi})$ and
$\Theta_{\mathcal{H}}(\mathcal{A}_{C}^{\Phi\otimes\Psi})$.  The
theorem in section 4 and the strong hereditary property of the sets
$\mathcal{A}_{E}^{\Phi\otimes\Psi}$ and
$\mathcal{A}_{C}^{\Phi\otimes\Psi}$ imply that all extreme points of
the sets $\Theta_{\mathcal{H}}(\mathcal{A}_{C}^{\Phi\otimes\Psi})$,
$\Theta_{\mathcal{H}}(\mathcal{A}_{E}^{\Phi\otimes\Psi})$,
$\Theta_{\mathcal{K}}(\mathcal{A}_{C}^{\Phi\otimes\Psi})$,
$\Theta_{\mathcal{K}}(\mathcal{A}_{E}^{\Phi\otimes\Psi})$ have the
same rank 1 and, hence, are product pure states. By propositions 5
and 6 all these pure states lie in the sets
$\mathcal{A}_{C}^{\Phi}$, $\mathcal{A}_{E}^{\Phi}$,
$\mathcal{A}_{C}^{\Psi}$, $\mathcal{A}_{E}^{\Psi}$ correspondingly.
This implies $"\subseteq"$ in the above projective
relations.$\square$

This proposition provides the following observation.

\textbf{Corollary 7.} \textit{Let the assumption B holds for the
channels $\Phi$ and $\Psi$. If the set
$\mathcal{A}_{E}^{\Phi\otimes\Psi}$
($\mathcal{A}_{C}^{\Phi\otimes\Psi}$) has entangled state $\omega$
then the states $\omega^{\mathcal{H}}$ and $\omega^{\mathcal{K}}$
can be represented as a convex combination of pure states from the
optimal sets $\mathcal{A}_{E}^{\Phi}$ and $\mathcal{A}_{E}^{\Psi}$
($\mathcal{A}_{C}^{\Phi}$ and $\mathcal{A}_{C}^{\Psi}$) for single
channels $\Phi$ and $\Psi$ correspondingly.}

The additivity of the minimal output entropy for two channels
$\Phi$ and $\Psi$ implies existence of nonentangled optimal states
for the channel $\Phi\otimes\Psi$, but does not forbid a
particular entangled state to be optimal (trivial example of
noiseless channels). Corollary 7 implies restriction on entangled
optimal states for the channel $\Phi\otimes\Psi$.  The partial
traces of any entangled optimal state must be represented as a
convex combination of optimal pure states for single channels
$\Phi$ and $\Psi$.

This implies the following observation. If one of the strongly
additive channels $\Phi$ and $\Psi$ has single optimal
state\footnote{the examples of channels with single optimal state
are considered in \cite{K-R}}, then the set of optimal states for
the channel $\Phi\otimes\Psi$ consists only of tensor products of
this single state with the optimal states of the another channel.
For example, let $\Phi$ be an arbitrary channel with single optimal
state and $\Psi$ be an identical or entanglement breaking channel
(or direct mixture of channels of this type), then we can assert
(due to proposition 2 in \cite{H-Sh-1}) that any optimal state for
the channel $\Phi\otimes\Psi$ has the above form of tensor products
and hence there is no entangled optimal state.

If the global additivity conjecture is true then every pair of
channels are strongly additive \cite{H-Sh-1},\cite{Sh-e-a-q}. So,
an example of two channels with entangled optimal state, whose
partial trace is not decomposable on optimal states of single
channels, implies the breaking of this conjecture.

\vskip10pt

\textbf{Acknowledgments.} The author is grateful to A. S. Holevo for
permanent support and to K. Audenaert for pointing out a flaw in the
initial version of this paper. The author also thanks M.B.Ruskai and
P.W.Shor for the help in searching the origin of the particular
results. The work was partially supported by INTAS grant 00-738.

\end{document}